\documentclass[12pt]{article}
\usepackage{epsfig,graphics,graphicx}

\begin{document}

\begin{center}
\large {\bf BIEXCITONIC LIQUID IN $\beta$-ZnP$_{2}$ CRYSTAL:
QUANTUM BOSE LIQUID AND ITS POSSIBLE SUPERFLUIDITY}

\bigskip

{\large I.\ S.\ Gorban\footnote{Deceased}, O.\ A.\ Yeshchenko, I.\
M.\ Dmitruk, and M.\ M.\ Bilyi}

\smallskip

{\it Physics Faculty, Kyiv Taras Shevchenko University, \\ Akad.
Glushkov pr. 6, 03127  Kyiv, Ukraine
\\E-mail: yes@mail.univ.kiev.ua}

\end{center}

\begin{abstract}
A phonon--roton dispersion relation is proposed for the elementary
excitations of a quantum biexcitonic liquid in semiconductors. The
proposed dispersion relation is used as a starting point for a
calculation of the photoluminescence spectrum of the liquid and an
analysis of its behavior  under variation of the temperature and
density of the biexcitonic liquid. The parameters of the
dispersion curve of elementary excitations of the quantum
biexcitonic liquid are evaluated by fitting the calculated
photoluminescence spectrum to the experimental spectrum of the
biexcitonic liquid of semiconducting $\beta $-ZnP$_{2}$ crystals.
Experimental studies of how the photoluminescence spectrum of a
biexcitonic liquid in $\beta $-ZnP$_{2}$ depends on the
temperature and the intensity of the laser excitation confirm the
initial theoretical model. The dependence of the temperature of
the crystals on the excitation intensity is measured, and for some
of the samples an anomalous dependence is found: the temperature
of the crystal decreases as the excitation intensity increases.
This effect is probably a consequence of the giant thermal
conductivity of the superfluid biexcitonic liquid in $\beta
$-ZnP$_{2}$ crystals.
\end{abstract}


dispersion relation, superfluidity

\smallskip


\newpage

\subsection*{\bf Introduction}

For a long time the phenomenon of superfluidity was observed
exclusively in liquid $^{4}$He at temperatures below 2.17 K. It is well
known that the dispersion relation for elementary excitations in superfluid He
II is characterized by a so-called roton gap. A dispersion relation of
this sort was first proposed by Landau\cite{1} and later investigated
theoretically by Feynman,\cite{2} and it was subsequently confirmed in neutron
scattering experiments.\cite{3} Among the Bose liquids, $^{4}$He is
unique in that it remains liquid at atmospheric pressure down to the
lowest temperatures. Other Bose liquids crystallize at temperatures
above those at which superfluidity can arise. Crystallization sets in
because the energy of the intermolecular interaction is greater than
the vibrational energy at temperatures above the temperature of the
onset of superfluidity. Since the vibrational energy goes as $\hbar
\omega _{\rm vibr}\propto m^{-1/2}$ ($m$ is the mass of the molecule),
superfluidity is more likely to arise in liquid molecular hydrogen
than in other liquids. However, even in the case of liquid H$_{2}$ the
energy of the intermolecular interaction is too large, and
crystallization sets in before superfluidity as the temperature is
lowered.

The analogy between Wannier--Mott excitons in semiconductors and
atomic hydrogen is well known. This analogy extends to excitonic
molecules (EMs or biexcitons) and to the hydrogen molecule. The
problem of Bose condensation and superfluidity in a system of
excitons has been very popular in semiconductor physics for the
past 20--25 years (see, e.g., Refs.\cite{4}-\cite{6}). The
possibility of superfluidity of excitons in a strong magnetic
field was pointed out in Ref.\cite{5}. An unusual ballistic
solitonlike regime of exciton motion in Cu$_{2}$O crystals was
observed in Ref.\cite{6}. This effect was interpreted as evidence
of superfluidity of Bose-condensed excitons.

Guided by the analogy between excitonic molecules and hydrogen
molecules, one can analyze the possibility for superfluidity to
arise in such a biexcitonic (molecular) liquid. The existence of a
biexcitonic liquid in semiconductors was first considered in
Ref.\cite{7}. As we have said, superfluidity is not observed in
liquid hydrogen because of the earlier onset of crystallization.
However, the effective mass of excitonic molecules is 2--3 orders
of magnitude smaller than that of the H$_{2}$ molecule, and
crystallization of a biexcitonic liquid probably should not occur
at all. Therefore, at sufficiently low temperatures a biexcitonic
liquid can go into a superfluid state. However, as it turns out,
the formation of a biexcitonic liquid itself is extremely
problematical. Keldysh\cite{8} and later Brinkman and Rice\cite{9}
pointed out two important differences between H$_{2}$ molecules
and EMs in typical semiconductors. First, since the effective
masses of the electron and hole are often of the same order, the
binding energy of the  excitonic molecule, measured in units of
the binding energy of the exciton (atom)  , should be much less
than 0.35 for the hydrogen molecule. Second, the contribution of
the energy of zero-point  motion is considerably larger in a
system of biexcitons because of the small mass of a biexciton. For
these two reasons, as experiments have shown, in typical
semiconductors (Si, Ge, and a number of others) at high excitation
intensities an electron--hole Fermi liquid is formed. One
therefore expects that the formation of a biexcitonic Bose liquid
can occur only in crystals in which the effective masses of the
electron and hole are  sufficiently different. Furthermore, the
mass of the EM should be sufficiently large that the contribution
of the zero-point  motion is sufficiently small and it becomes
possible for a molecular liquid to form. At the same time, the
mass of the EM must be small enough that ``early'' crystallization
cannot occur.

Crystals of monoclinic zinc diphosphide ($\beta $-ZnP$_{2}$) meet
the stated requirements. This crystal is characterized by an
appreciable difference of the electron and hole effective masses
($\sigma  = m_{e}/m_{h} = 0.06$), a rather large value of the
translational mass of the biexciton $m_{\rm bex} = 3m_0$ ($m_0$ is
the free electron mass), and a very strong anisotropy of the
effective masses of the carriers. For example, the anisotropy
parameter $\gamma  = \mu _{\parallel}/\mu _{\perp }$ is equal to
0.19 ($\mu_{\parallel }$ and $\mu _{\perp }$ are the values of the
reduced masses of an exciton in the directions parallel to and
perpendicular to the $Z$ ($c)$ axis of the crystal). As we know,
the anisotropy leads to an increase in the binding energy of free
electron--hole complexes, i.e., excitons, biexcitons, and their
condensates. Therefore, the excitonic molecule in $\beta
$-ZnP$_{2}$ has a  rather high binding energy $E_{\rm bex}^{b} =
6.7\ {\rm meV} = 0.15E_{\rm ex}^{b}$.

The excitonic state with the lowest energy in this crystal is the
dipole-forbidden $1S$ state of the orthoexciton.\cite{10} This
makes it rather easy to create an appreciable concentration of
excitons and, hence, biexcitons under laser excitation. Studies by
the authors have shown that the condensation of biexcitons in
$\beta $-ZnP$_{2}$ crystals occurs via a hydrogenlike scenario,
i.e., unlike the case of typical semiconductors, a liquid of the
molecular (insulator) type forms in them. In the photoluminescence
(PL) spectrum of these crystals one observes the so-called $C$
line, which is due to two-photon annihilation of
biexcitions\footnote{ At two-photon annihilation there is no
recoil particle that could  take an appreciable  part of the
momentum of the annihilating EM. For this reason, since the photon
momentum is very small, momentum conservation allows two-photon
decay only for an EM with $k\approx 0$, i.e., the two-photon PL
line should be very narrow.\cite{14} In two-electron (one-photon)
transitions, in which only one exciton of a molecule annihilates,
the second, surviving exciton can  take an arbitrary
quasimomentum, i.e., the one-photon PL lines are rather wide. This
makes it possible to distinguish the two-photon PL lines of free
biexcitons and of a biexcitonic liquid, unlike the case of the
one-photon PL lines.} condensed into a molecular  (insulator)
liquid.\cite{11,12,13} In Refs.\cite{12} and \cite{13} the phase
diagram of a biexcitonic liquid was measured and its critical
parameters were determined: $T_{C} = 4.9$ K, $n_{{\rm bex},C} =
4.1\times 10^{18}$ cm$^{-3}$ ($r_{{\rm bex},C}\approx 63\ {\rm
\AA} = 4.2a_{\rm ex}$, where $a_{\rm ex}$ is the excitonic Bohr
radius).

In this paper we propose a model for the quantum biexcitonic Bose
liquid and calculate its emission spectrum. In the framework of
this model we analyze the experimentally  observed features of the
fine structure of the emission line of a biexcitonic liquid in
$\beta $-ZnP$_{2}$ (the $C$ line) as the temperature and the
intensity of the laser excitation of the crystal are varied.

\subsection*{Dispersion relation for elementary excitations of a
quantum biexcitonic liquid}

Thus we assume that the biexcitonic liquid does not crystallize down to
the temperatures at which quantum effects become important. What sorts
of elementary excitations can exist in a quantum biexcitonic Bose
liquid? It is logical to assume that, first, there are acoustic
phonons. Their dispersion relation is given as
\begin{eqnarray}
E(k) = \hbar uk  \, ,
\label{1}
\end{eqnarray}
where $u$ is the  sound velocity, and $k$ is the wave vector.
Second, we assume that owing to the intermolecular interaction in
the liquid a collectivization of the intramolecular excitations
can occur in it. Since the lowest-energy excited state of an
excitonic molecule is a rotational state,\cite{15} these
collectivized molecular excitations can be rotational excitations
of the molecules. Here, following Landau,\cite{1} these
collectivized excitations/quasiparticles will be called rotons.
Rotons in a biexcitonic (molecular) liquid can be regarded as
molecular Frenkel excitons. The excitation energy of the molecule
in the presence of phase correlation between the molecules in the
liquid can be written as
\begin{eqnarray}
\Delta E = \Delta E_{0} + D + 2M\cos [(k - k_{0}^{\prime})a)] \, ,
\label{2}
\end{eqnarray}
where $\Delta E_0$ is the excitation energy of a free molecule,
$D$ is the change of the interaction energy of a given molecule
with its neighbors  under its excitation, $M$ is the matrix
element for the transfer of the energy of excitation from the
excited molecule to a neighbor  being in the ground state, $a$ is
the  average distance between these molecules, and $k$ is the wave
vector of a Frenkel exciton, i.e., in our case the roton. The last
term in Eq.\ (2) describes the dispersion relation of rotons and
is physically meaningful for $|\Delta k| = |k-k_0'| \leq \pi /2a$.
Thus the photon--roton dispersion relation for elementary
excitations of a biexcitonic liquid can be written as
\begin{eqnarray}
E(k) = \hbar uk \; {\rm at} \; 0 \leq k < k_{0}^{\prime} - \pi/2a
\\ \nonumber {\rm and} \; k > k_{0}^{\prime} + \pi/2a \, ; \\
\nonumber E(k) = \hbar uk - \gamma \cos [(k - k_{0}^{\prime})a)]
\\ \nonumber {\rm at} \; k_{0}^{\prime} - \pi/2a \leq k \leq
k_{0}^{\prime} + \pi/2a \, , \label{3}
\end{eqnarray}
where $\gamma  = -2M$. For small $|\Delta k|$ the second relation of
system (3) can be written in the form
\begin{eqnarray}
E_{r}(k) = \Delta + \frac{\hbar ^{2} (k-k_{0})^{2}}{2m_{r}} \, .
\label{4}
\end{eqnarray}
Equation (4) is exactly the same as the equation proposed by
Landau\cite{1} for describing the dispersion relation of rotons in
superfluid He II. Thus our choice of the term ``roton'' is not
arbitrary but is based on the similarity of the dispersion relations
of rotons in helium and the collectivized molecular excitations of a
biexcitonic liquid. In Eq.\ (4) we have introduced the following
notation: $m_{r} = -\hbar ^{2}/2Ma^{2}$ is the roton effective mass,
$k_0 = k_0'-m_{r}u/\hbar $ is the wave vector corresponding to the
roton minimum of the dispersion relation, and $\Delta  = \hbar
uk_0+2M+m_{r}u^{2}/2$ is the energy gap in the spectrum of elementary
excitations of the biexcitonic liquid.

Starting from the dispersion relation (3) for a quantum
biexcitonic liquid, let us calculate the PL spectrum for such a
liquid. The spectrum we are looking for can be obtained by
convolution of the slit function and the function $I(E) = I_0\rho
(E)f(E)$, where $I_0 = {\rm const}$, $\rho (E) = \rho
_0k^{2}(E)(dk(E)/dE)$ is the density of states of the liquid,
which is determined from the dispersion relation for the
elementary excitations ($\rho _0 = {\rm const}$); $f(E) =
1/[\exp(E/k_{B}T)-1]$ is the distribution function for the
excitations/quasiparticles of the liquid (the Bose--Einstein
distribution function). The zero of energy $E$ is taken as the
value corresponding to the radiative transition from the state
with $k = 0$. For the parameters of the dispersion curve we used
the values obtained from a best fit of the calculated luminescence
spectrum to the experimental PL spectrum of a biexcitonic liquid,
recorded for a $\beta $-ZnP$_{2}$ crystal of high purity. Figures
1, 2, 3, and 4 show the calculated PL spectra of a quantum
biexcitonic liquid of various densities at different temperatures
and also the characteristics of these spectra as functions of
temperature $T$ and the square of the excitation intensity $I_{\rm
exc}^{2}$ (the density of the biexcitonic liquid is proportional
to $I_{\rm exc}^{2}$). It is seen that the PL spectrum of a
biexcitonic liquid should have a two-component structure. In the
proposed model it is assumed that the shape and parameters of the
dispersion curve  do not depend on temperature and depend only  on
the density of the liquid . The density of the liquid influences
the shape of the dispersion curve and, through it, the shape of
the PL spectrum. We used the following proportionalities relating
the parameters of the dispersion curve and the density of the
liquid (which is proportional to the square of the excitation
intensity): average intermolecular distance $a\propto
n^{-1/3}\propto I_{\rm ex}^{-2/3}$,  sound velocity $u\propto
n^{1/2}\propto I_{\rm exc}$, $k_0'a = {\rm const}$, and the
modulus of the matrix element for the excitation transfer between
molecules of the liquid $|M|\propto a^{-3}\propto I_{\rm
exc}^{2}$.

Concluding this Section we must mention the following. A
short-range order should be established in the system of
biexcitons  due to the substantial interaction between them;
however, there is apparently no long-range order. The presence of
short-range order ensures that relation (2) will apply at least in
a qualitative way.

\subsection*{Experimental details}

We studied semiconducting single crystals of monoclinic zinc
diphosphide of high purity, grown from the gas phase. The
technology of the synthesis and growth of these crystals is
described in Ref.\cite{16}. For excitation of the luminescence we
used a cw Ar$^{+}$ laser (emission line 5145 \AA{}).

During laser excitation the temperature of the crystal at the
point of excitation differs substantially from the temperature of
the external environment, i.e., of the helium bath surrounding the
sample and the unilluminated region of the sample. This makes it
impossible to use a thermosensor for precision temperature
measurements, as one would get some averaged temperature. It was
therefore necessary to use an internal temperature standard (ITS)
that would make it possible to determine the temperature of the
crystal at the point of excitation. For this standard we used the
temperature dependence of the spectral position of the narrow,
intense $B$ line of the PL spectrum (the emission line of the free
forbidden orthoexciton). Because of the  variation of the energy
gap  $E_{g}$ with  variation of temperature, the $B$ line changes
its spectral position (shifts to lower energy with increasing
temperature). Since the shift of the B line with temperature is
small ($dT/d\lambda _{B} = 3.625$ K/\AA), for correct measurement
we recorded a reference line of neon in the investigated spectral
region simultaneously with the PL spectrum. This technique made it
possible to determine the temperature of the sample at the point
of excitation to an accuracy of 0.05 K.

\subsection*{Biexcitonic liquid in $\beta $-ZnP$_{2}$
crystals. Experimental results and discussion}

At excitation intensities above 1 kW/cm$^{2}$ and temperatures
below 5 K the PL spectrum of $\beta $-ZnP$_{2}$ crystals, as we
have said, contains an emission line of the biexcitonic liquid
(the $C$ line, henceforth called the $C$ spectrum). As we see from
Figs.\ 5, 6, 7, 8, and 9, the $C$ spectrum has a two-component
form, as was predicted by the model set forth above. Using the
dispersion relation of elementary excitations of a biexcitonic
liquid (3)  we fit the experimental PL spectrum of the biexcitonic
liquid (Fig.\ 5), which was obtained for a sample of high purity
(excitation energy 3 kW/cm$^{2}$). This made it possible to
determine the parameters of the dispersion curve of the elementary
excitations: $\Delta  = 0.49$ meV, $k_0 = 9.8\times 10^{6}$
cm$^{-1}$, $u = 1.4\times 10^{5}$ cm/s, and $m_{r} = 2.2m_0$
($m_0$ is the free electron mass), and also the value of the
intermolecular distance in the liquid $a = 92\ {\rm \AA} =
6.1a_{\rm ex}$ and the  temperature of the crystal at the point of
excitation, $T = 1.5$ K. The values obtained are physically
reasonable, a fact which, in our view, tends to confirm that the
proposed theoretical model is correct. Further evidence of this is
the good agreement of the experimental and theoretical spectra, in
view of the approximate, model character of the dispersion curve
given by Eqs.\ (3).

The evolution of the shape of the $C$ spectrum and the behavior of
its characteristics  under variation of temperature and  of the
square of the excitation intensity are shown in Figs.\ 6, 7, 8,
and 9. The behavior of the experimental spectrum under variation
of temperature (Figs.\ 6 and 7) is similar to that of the
theoretical spectrum (Figs.\ 1 and 2). Analysis of the $C$
spectrum  under variation of the excitation intensity is made
difficult by the fact that simultaneously with the  variation of
excitation intensity there is also a  variation of temperature,
which has a definite effect on the shape of the spectrum. In
comparing Figs.\ 2, 4, and 9 we can conclude that the change in
the $C$ spectrum at moderate levels of excitation is probably due
mainly to the influence of temperature (the rise in temperature
with increasing $I_{\rm exc}$), while at higher excitation
intensities the evolution of the shape of the $C$ spectrum occurs
mainly on account of the increase in the density of the liquid
with increasing $I_{\rm exc}$. What we have said agrees with the
intensity dependence of the temperature of the crystal at the
point of excitation, shown in Fig.\ 10. Thus the experiment is
described by the theory in a completely satisfactory way. The
shape of the $C$ spectrum can differ substantially for different
samples. This is due to the fact that at the same values of the
excitation intensity and crystal temperature, a liquid of higher
density should arise in purer samples than in less pure samples.
Consequently, the shape of the PL spectrum of the liquid varies
with its density.

The results presented above suggest that a quantum biexcitonic
liquid characterized by a phonon--roton dispersion relation for
the elementary excitations can form in $\beta $-ZnP$_{2}$ crystals
under certain conditions. This sort of dispersion relation, as we
know, is evidence that superfluidity can arise in a Bose liquid,
in particular, in liquid $^{4}$He. Is it possible that
superfluidity can arise in a biexcitonic liquid in a $\beta
$-ZnP$_{2}$ crystal? To answer this question we  performed the
following experiments. For some of the samples (Fig.\ 10) we
studied the temperature of the crystal at the point of excitation
as a function of the intensity of the excitation. The normal
situation is for the temperature of the crystal to increasing with
increasing excitation intensity. Besides the normal monotonically
increases behavior of $T(I_{\rm ex}^{2})$ (curve {\em 1}), for
several samples we obtained anomalous $T(I_{\rm ex}^{2})$ curves:
decreasing (curve {\em 2}) or nonmonotonic (curve {\em 3}). We
think that the anomalous $T(I_{\rm ex}^{2})$  dependences can be
explained by an anomalously large (giant) thermal conductivity,
which accompanies the appearance of the superfluid ($s)$ component
in a quantum liquid below the superfluid transition. Such an
effect is well known for superfluid He II. We propose the
following explanation for the anomalous behaviors observed. At a
certain excitation intensity the density of the liquid reaches
values sufficient for a transition of the liquid to the superfluid
state. Upon further increase in $I_{\rm exc}$ the density of the
liquid increases, with the result that the temperature of the
superfluid transition increases and, with it, the fraction  of the
$s$ component. This can increase the thermal conductivity, i.e.,
the thermal energy should be removed more efficiently from the
excitation region of the crystal, and so the temperature of the
crystal should decrease with increasing excitation intensity.
Apparently those crystals with a monotonically decreasing
$T(I_{\rm ex}^{2})$ curve (curve {\em 2} in Fig.\ 10) are the
purest and most perfect: they have a low concentration of lattice
defects, and therefore a biexcitonic liquid is formed in them with
a sufficiently high density for the onset of superfluidity and the
corresponding giant thermal conductivity. Those crystals with a
nonmonotonic $T(I_{\rm ex}^{2})$ curve (curve {\em 3} in Fig.\ 10)
would be less pure. The liquid formed in them is of a lower
density but is nevertheless sufficient for the appearance of the
$s$ component. However this component represents a smaller
fraction than in the perfect crystals, and the giant thermal
conductivity  provides a sufficiently effective heat removal only
up to certain values of the  incoming laser power. When the power
is increased further, the temperature of the crystal begins to
grow. In those crystals with a monotonically increasing $T(I_{\rm
ex}^{2})$ dependence (curve {\em 1} in Fig.\ 10) the concentration
of defects is rather high, and therefore the density of the liquid
in them does not reach the values necessary for a transition to
the superfluid state and the resulting onset of giant thermal
conductivity. As a result, the temperature of the crystal at the
point of excitation increases with increasing excitation
intensity.

This study was supported in part by the Foundation for Basic Research of
the Ministry of Scientific and Technological Affairs of Ukraine
(Project No. 2.4/311).

\newpage

\begin{center}

\end{center}

\newpage

\begin{center}
\large{\bf Figure captions}
\end{center}

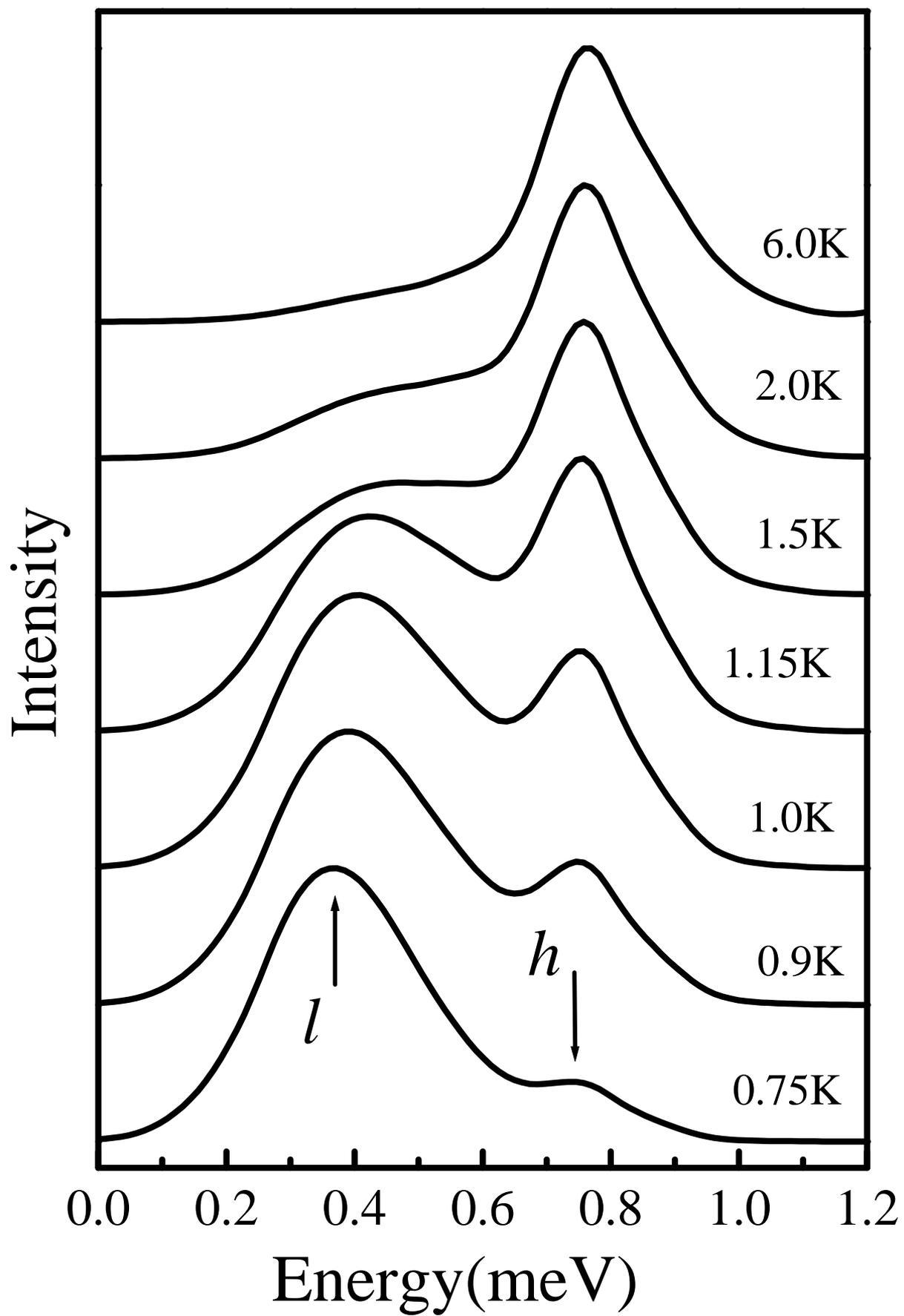
{\bf Figure1.} Evolution with temperature of the calculated
photoluminescence spectrum of a quantum biexcitonic liquid with a
phonon--roton dispersion relation for the elementary excitations.
The arrow labeled $l$ indicates the low-energy ($l)$ component of
the spectrum, and the arrow $h$ the high-energy ($h)$ component.

\bigskip

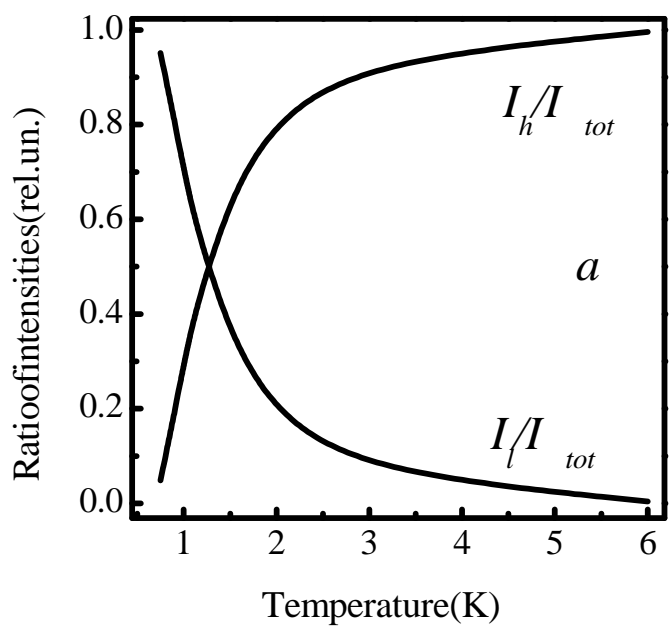
{\bf Figure2.} Temperature dependence of the characteristics of
the theoretical PL spectrum of a quantum biexcitonic liquid:
$I_{l}/I_{\rm tot}$ is the ratio of the intensity of the $l$
component to the total intensity of the $C$ spectrum,
$I_{h}/I_{\rm tot}$ is the same for the $h$ component (a);
$I_{l}/I_{h}$ is the ratio of the intensities of the $l$ and $h$
components (b); the positions of the maxima of the components (c);
the spectral distance (splitting) between the maxima of the
components (d); the half-widths of the components (e) and (f).

\bigskip

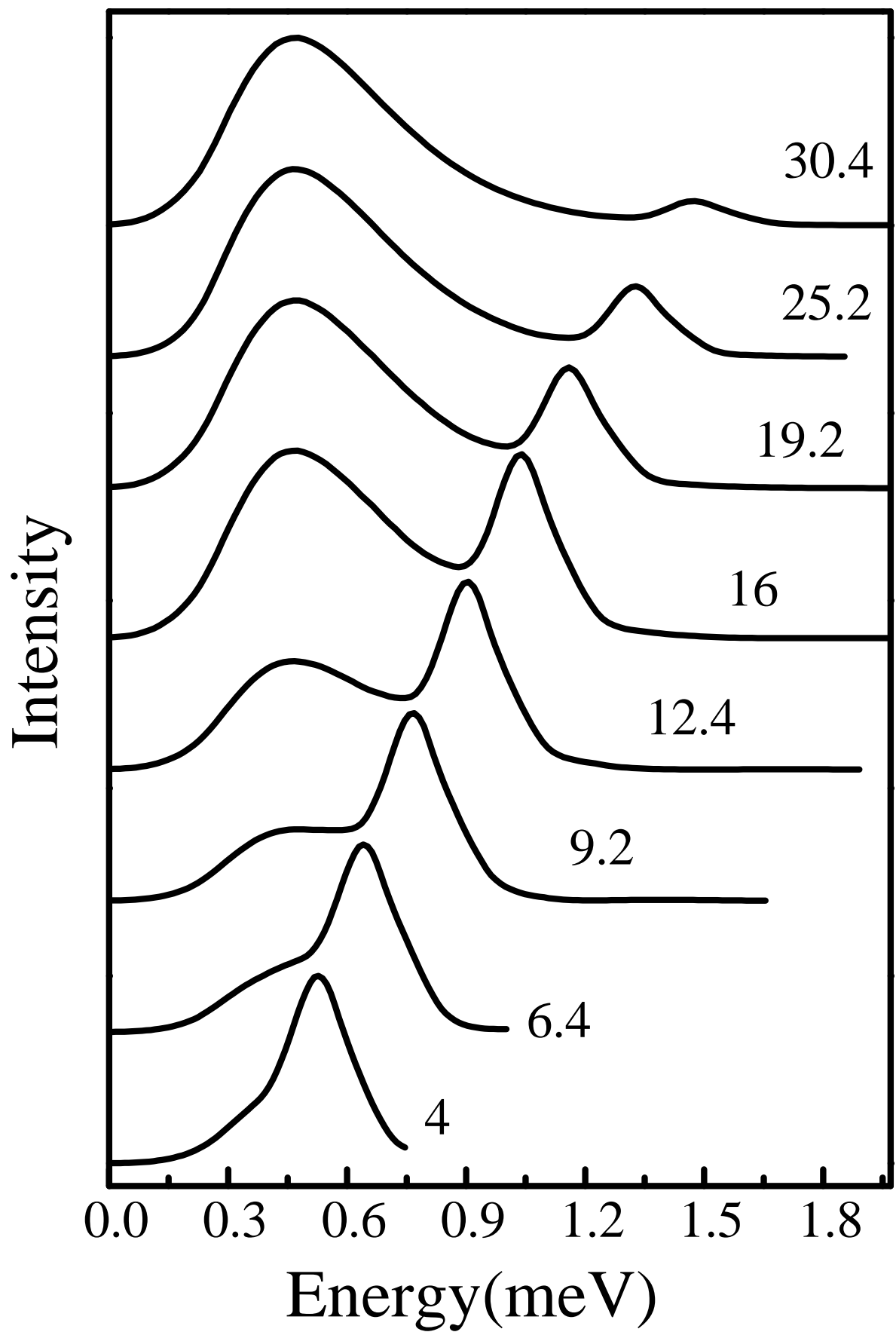
{\bf Figure3.} Evolution of the theoretical PL spectrum of a
quantum biexcitonic liquid  under variation of the square of the
excitation intensity (density of the liquid): $n\propto I_{\rm
ex}^{2}$. The curves are labeled with the value of $I_{\rm
ex}^{2}$ in arbitrary units.

\bigskip

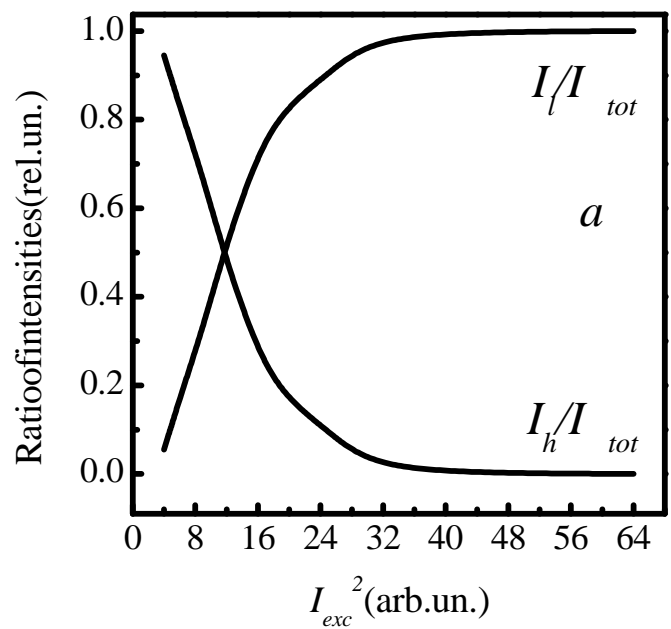
{\bf Figure4.} Characteristics of the theoretical PL spectrum of a
quantum biexcitonic liquid as functions of the square of the
excitation intensity (density of the liquid). The notation is the
same as in Fig.\ 2.

\bigskip

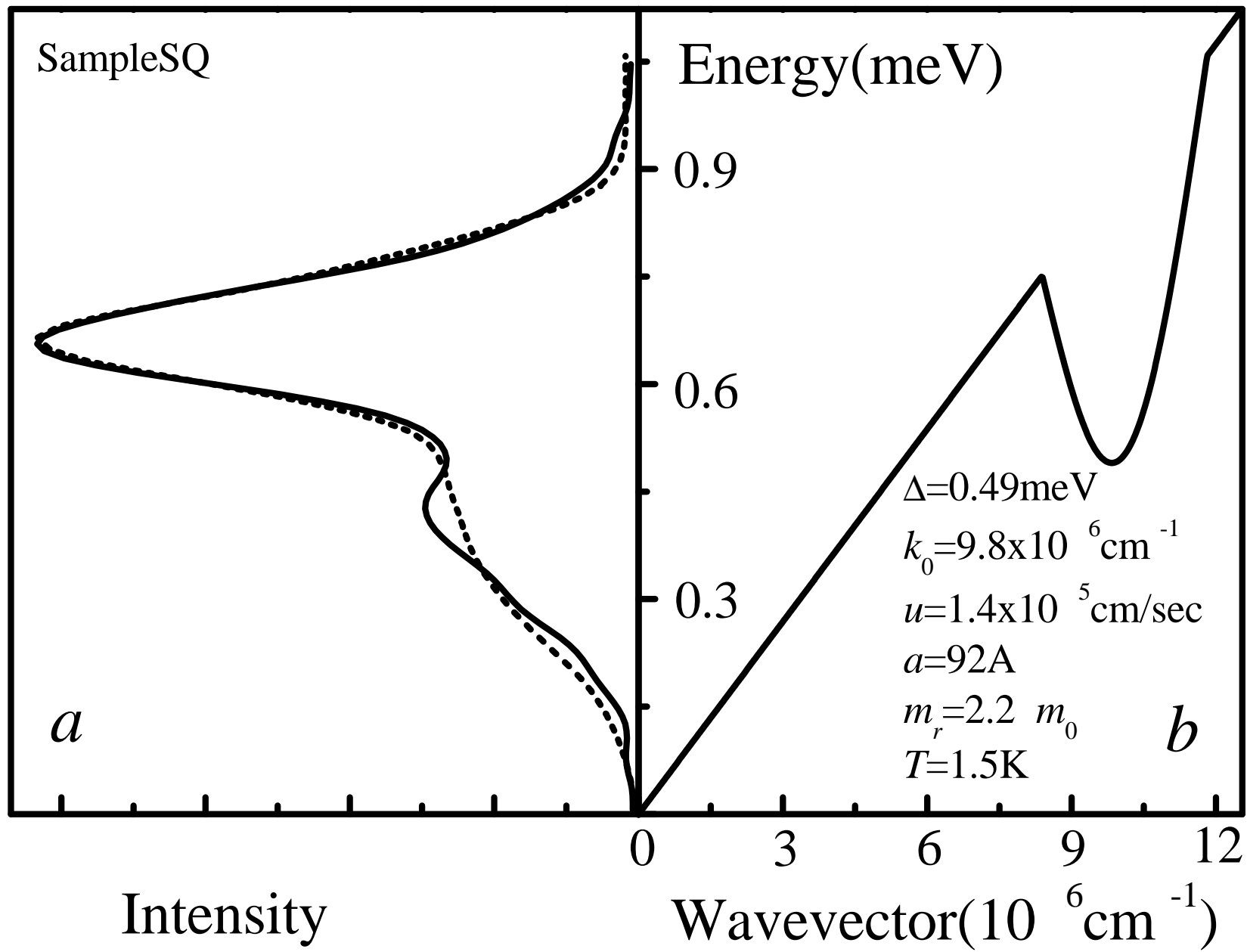
{\bf Figure5.} Experimental PL spectrum of a biexcitonic liquid in
a $\beta $-ZnP$_{2}$ crystal (solid curve) and its approximation
by the theoretical spectrum (dashed curve) (a); the dispersion
curve for the elementary excitations of the quantum biexcitonic
liquid with the parameters obtained by fitting the theoretical PL
spectrum to the experimental spectrum (b). Excitation intensity 3
kW/cm$^{2}$.

\bigskip

{\bf Figure6.} Evolution of the experimental PL spectrum of a
biexcitonic liquid in a $\beta $-ZnP$_{2}$ crystal  under
variation of temperature of the sample at the point of excitation.
Excitation intensity 4.8 kW/cm$^{2}$.

\bigskip

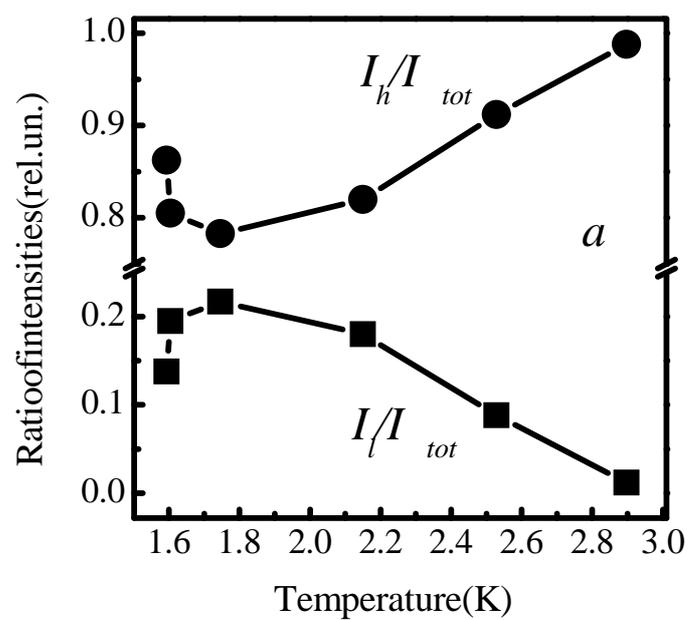
{\bf Figure7.} Temperature dependence of the characteristics of
the experimental PL spectrum of a biexcitonic liquid in a $\beta
$-ZnP$_{2}$ crystal. The notation is the same as in Figs.\ 2 and
4. Plotted along the vertical axis in part Fig.\ 7c is the
difference $[E_{m}(T)-E_{b}(T)]-[E_{m}(T_0)-E_{b}(T_0)]$, where
$E_{m}$ is the position of the maximum of the $l$ or $h$
component, and $E_{b}$ is the position of the maximum of the
narrow line of the forbidden free orthoexciton ($B$ line), which
was used to determine the temperature of the crystal at the point
of excitation, and $T_0$ is the minimum temperature  reached in
the experiment.

\bigskip

{\bf Figure8.} Evolution of the experimental PL spectrum of a
biexcitonic liquid in a $\beta $-ZnP$_{2}$ crystal under variation
of the square of the excitation intensity. The curves are labeled
by the value of $I_{\rm exc}^{2}$ in (kW/cm$^{2}$)$^{2}$.

\bigskip

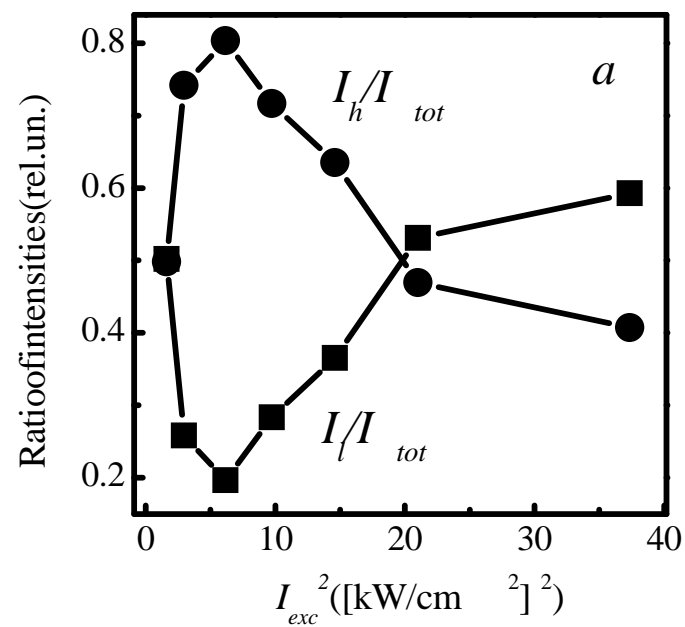
{\bf Figure9.} Characteristics of the experimental PL spectrum of
a biexcitonic liquid in a $\beta $-ZnP$_{2}$ crystal under
variation of the square of the excitation intensity. The notation
is the same as in Figs.\ 2 and 4 and 7. Plotted along the vertical
axis in Fig.\ 9c is the difference $[E_{m}(I_{\rm
ex}^{2})-E_{b}(I_{\rm ex}^{2})]-[E_{m}(I_0^{2})-E_{b}(I_0^{2})]$,
where $E_{m}$ is the position of the maximum of the $l$ or $h$
component, $E_{b}$ is the position of the maximum of the $B$ line,
and $I_0$ is the lowest excitation intensity used in the
experiment.

\bigskip

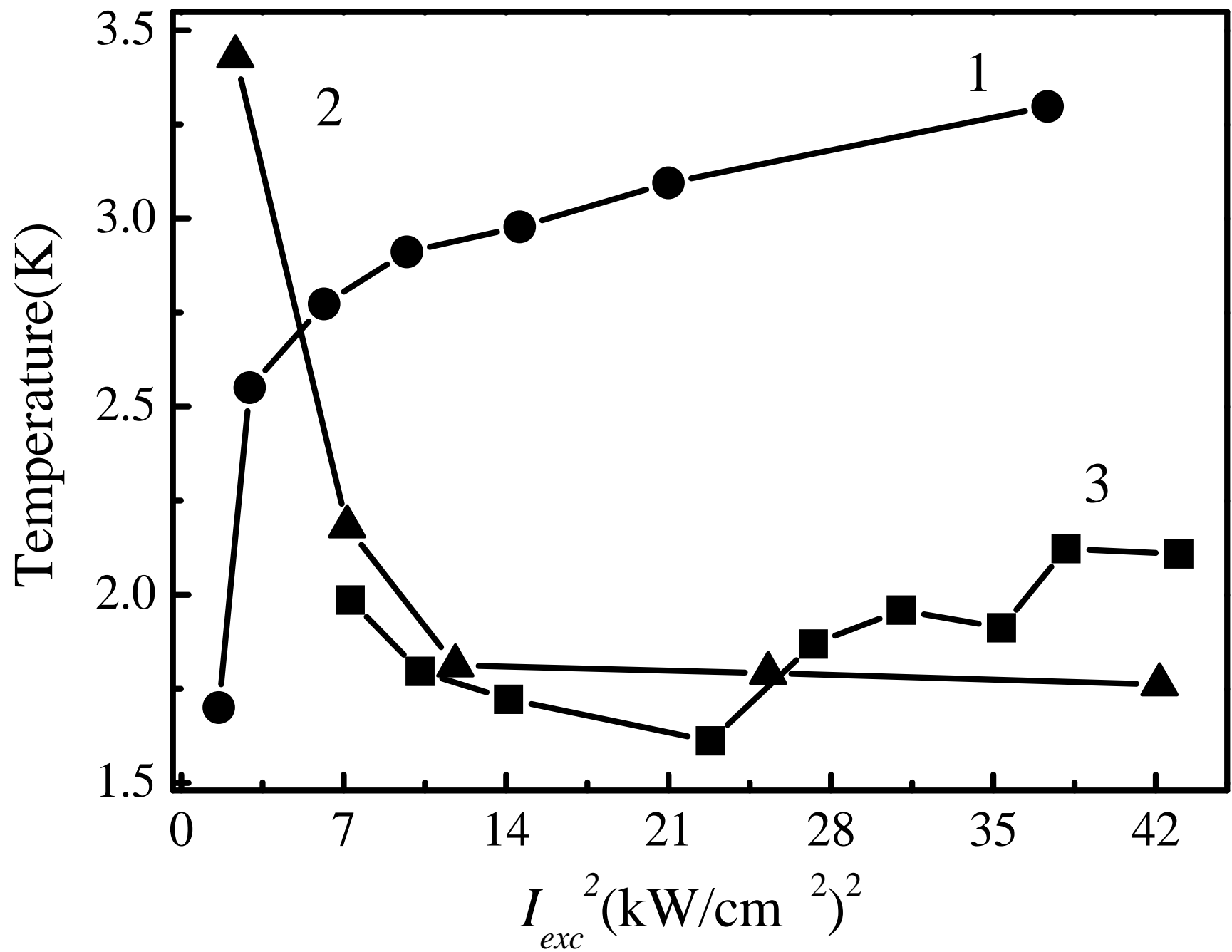
{\bf Figure10.} Temperature of the crystal versus the square of
the excitation energy for different samples: ZJ09-9 ({\em 1}),
ZJ09-14 ({\em 2}), and ZD99-1 ({\em 3}).

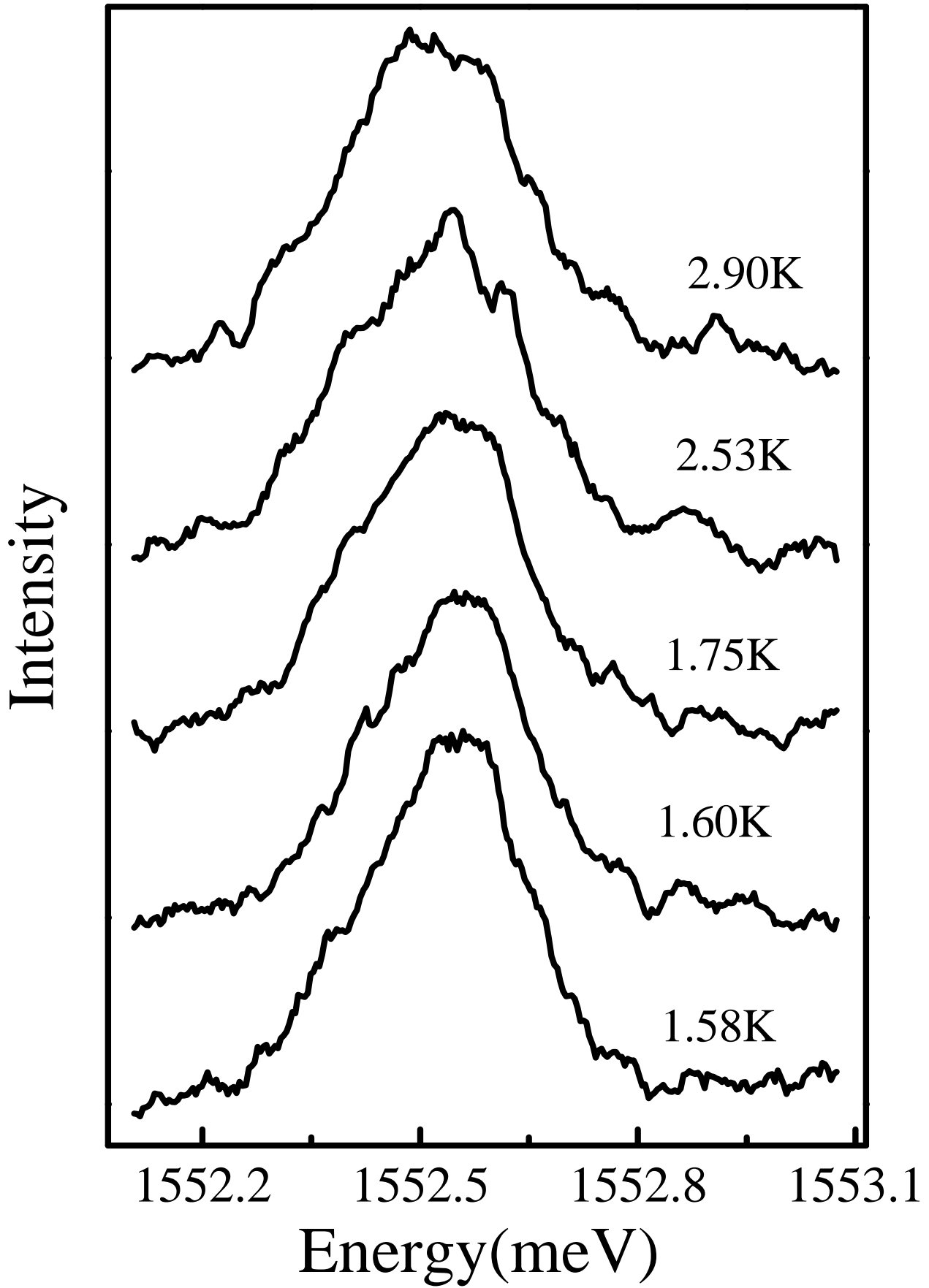
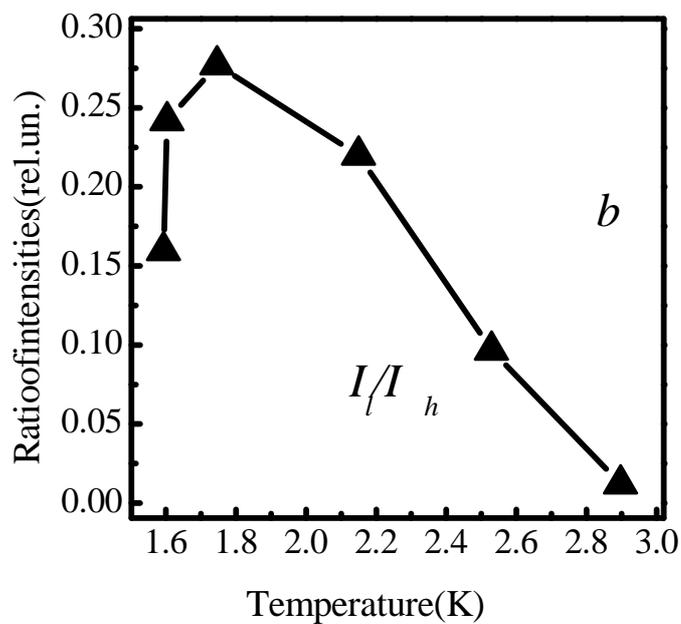
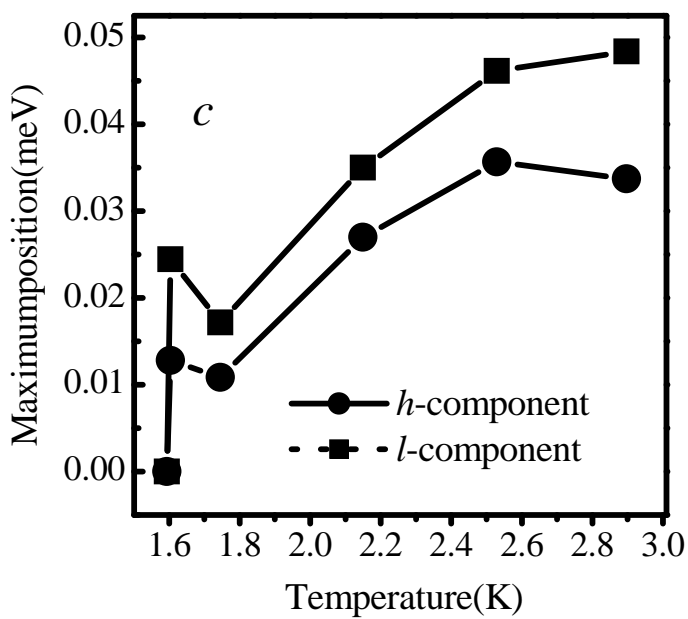
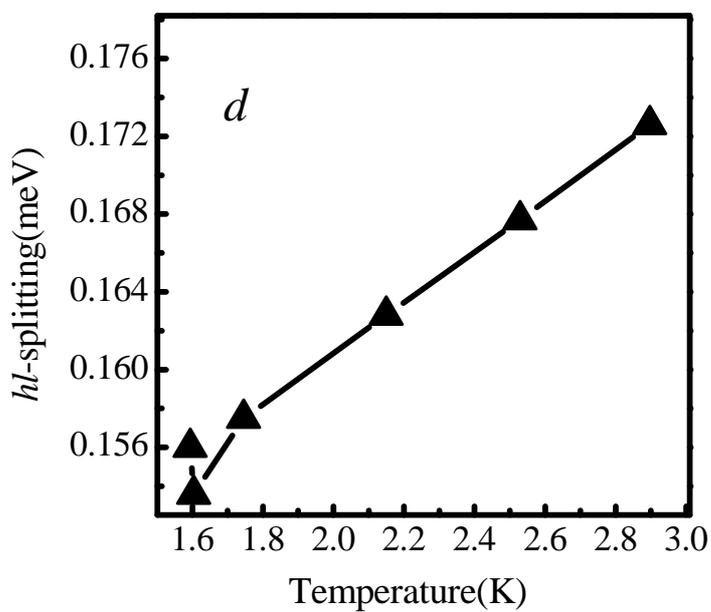
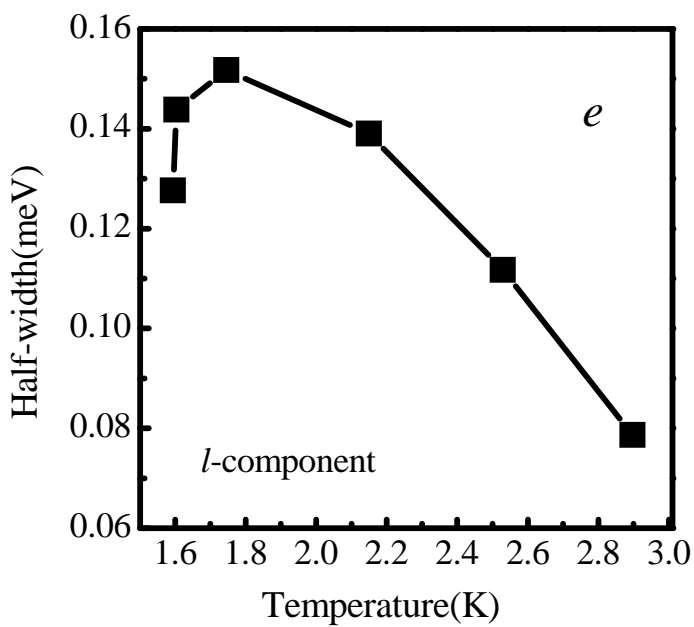
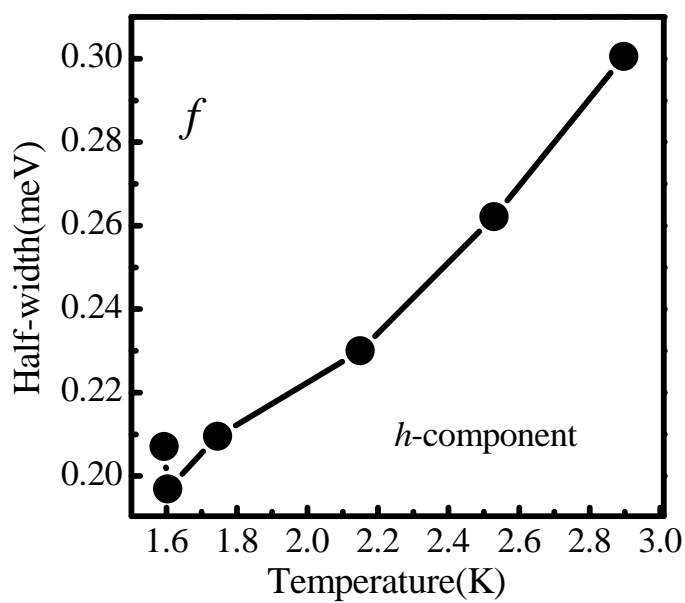
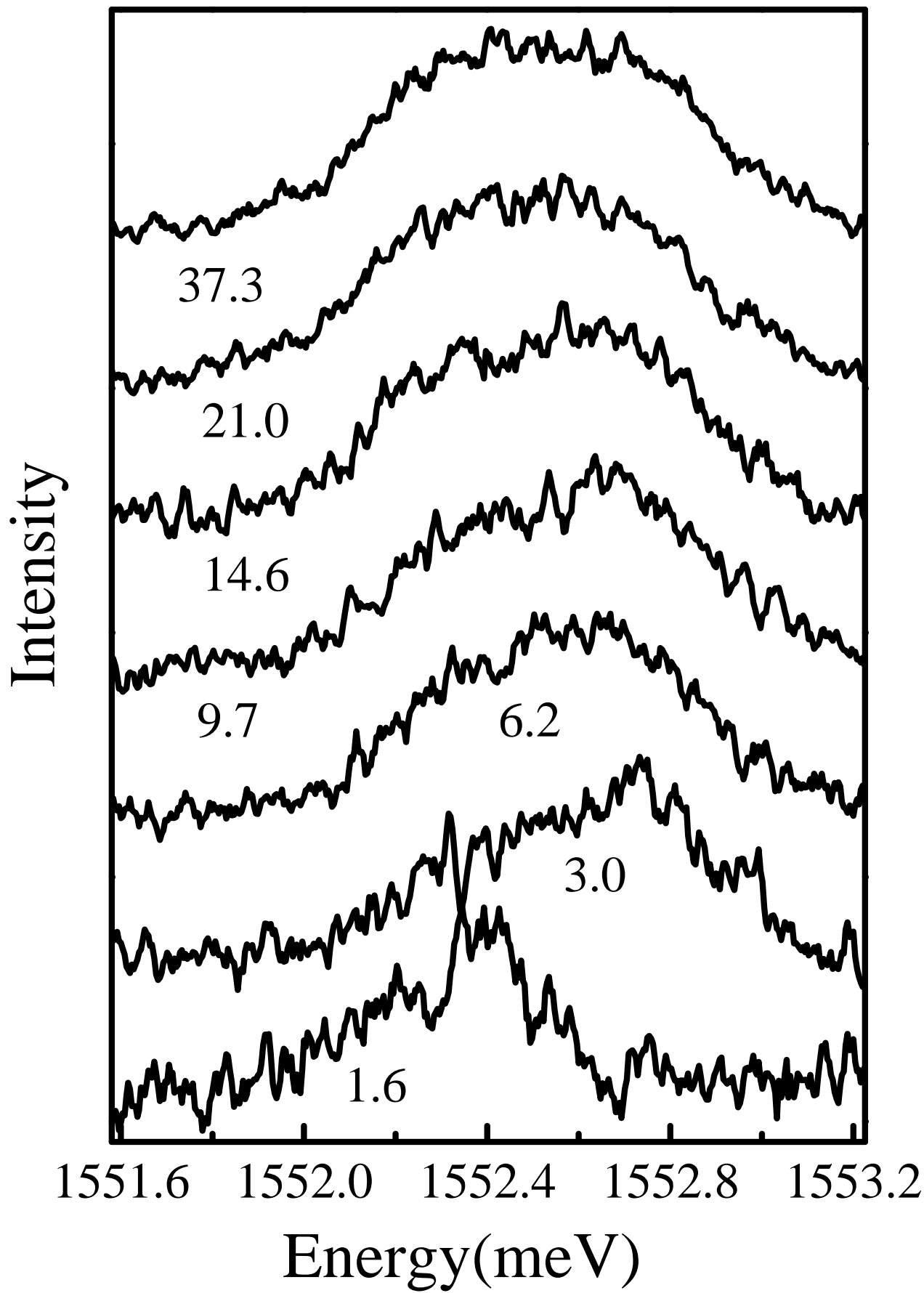
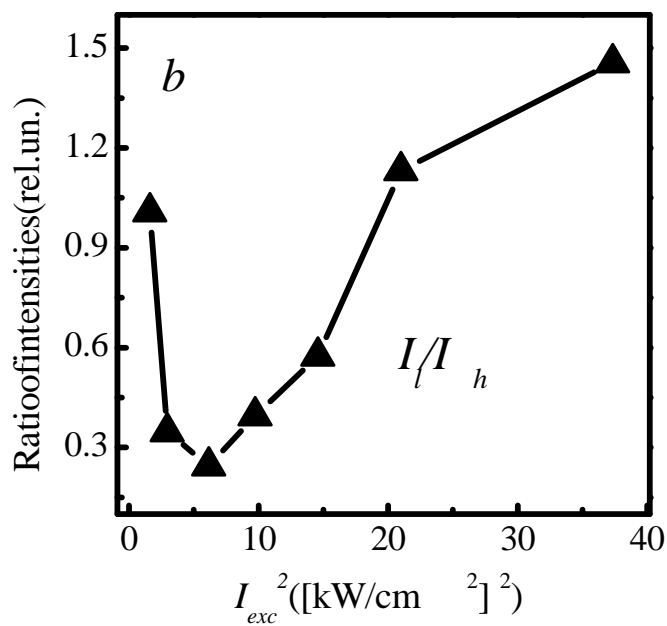
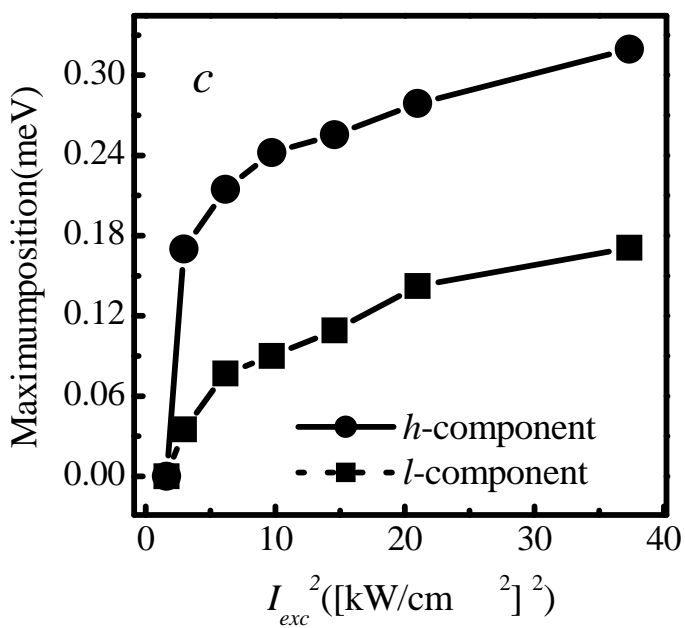
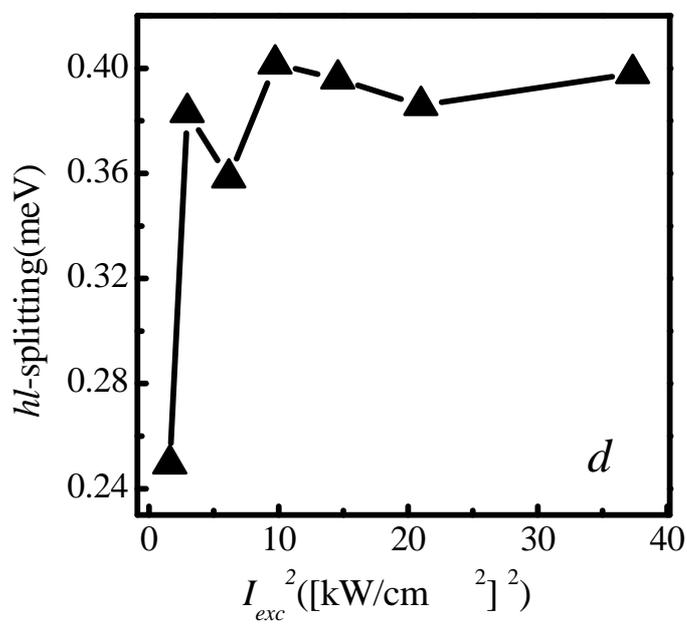
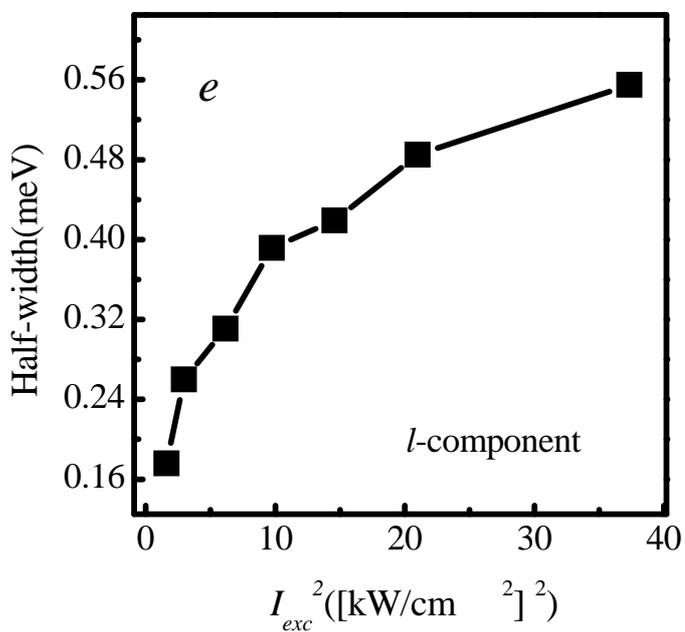
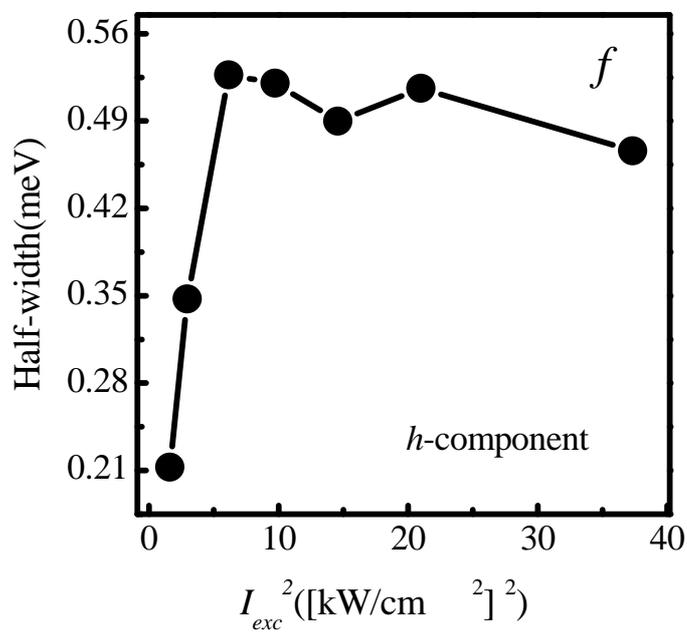
\end{document}